\documentclass[sigconf]{acmart}

\usepackage{balance}       
\usepackage{graphics}      
\usepackage{color}
\usepackage{booktabs}
\usepackage{paralist}
\usepackage{enumitem}
\usepackage{csquotes}
\usepackage{booktabs} 
\usepackage{etoolbox} 
\newtoggle{anonymous} 

\graphicspath {{figures/}}

\setcopyright{rightsretained}

\acmDOI{10.475/123_4}

\acmISBN{123-4567-24-567/08/06}

\iftoggle{anonymous}{
\acmConference[In submission to SIGIR'18]{TBD}{July 2018}{Ann Arbor, Michigan, USA} 
}{
\acmConference[Draft]{}{}{} 
}
\acmYear{2018}
\copyrightyear{2018}

\acmArticle{4}
\acmPrice{15.00}

\editor{Editor 1}
\editor{Editor 2}
\editor{Editor 3}

\begin{document}
\title{Improving Recommender Systems Beyond the Algorithm}

\iftoggle{anonymous}{
\author{Author names omitted}
\affiliation{%
  \institution{~}
  \city{~} 
  \state{~}
  \country{}}
\email{ }
}{
\author{Tobias Schnabel}
\affiliation{%
  \institution{Cornell University}
  \city{Ithaca} 
  \state{NY}
  \country{USA}}
\email{tbs49@cornell.edu}

\author{Paul N.~Bennett}
\affiliation{%
  \institution{Microsoft Research}
  \city{Redmond} 
  \state{WA}
  \country{USA}}
\email{pauben@microsoft.com}

\author{Thorsten Joachims}
\affiliation{%
  \institution{Cornell University}
  \city{Ithaca} 
  \state{NY}
  \country{USA}}
\email{tj@cs.cornell.edu}

}

\begin{abstract}
Recommender systems rely heavily on the predictive accuracy of the learning algorithm. Most work on improving accuracy has focused on the learning algorithm itself. We argue that this algorithmic focus is myopic. In particular, since learning algorithms generally improve with more and better data, we propose shaping the feedback generation process as an alternate and complementary route to improving accuracy. To this effect, we explore how changes to the user interface can impact the quality and quantity of feedback data -- and therefore the learning accuracy. Motivated by information foraging theory, we study how feedback quality and quantity are influenced by interface design choices along two axes: information scent and information access cost. We present a user study of these interface factors for the common task of picking a movie to watch, showing that these factors can effectively shape and improve the implicit feedback data that is generated while maintaining the user experience.
\end{abstract}

%
%



\maketitle

\section{Introduction}
\begin{figure}
\centering
  \includegraphics[width=0.95\columnwidth]{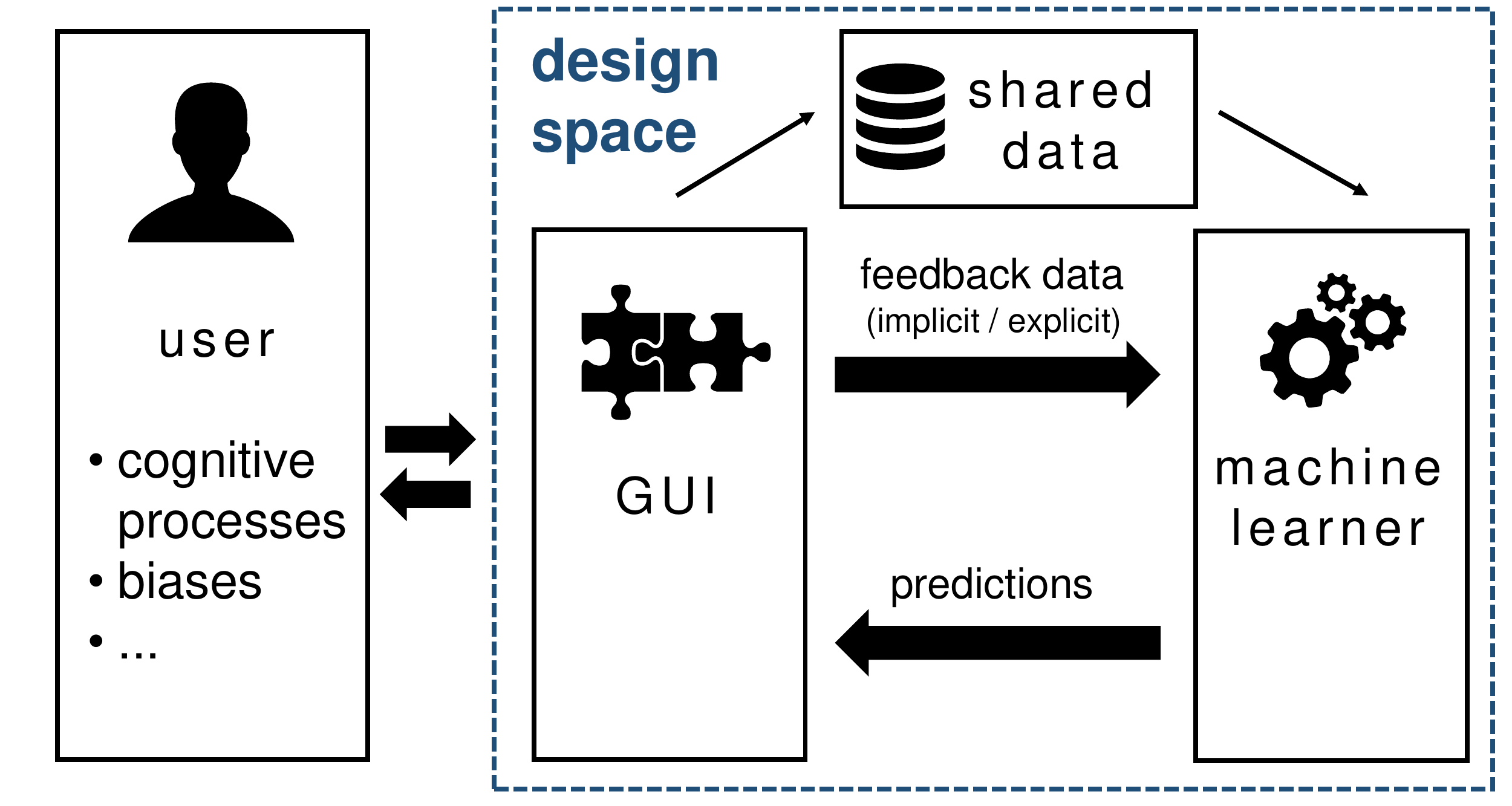}
  \caption{Typical recommender system. Many components are dependent on each other, providing a rich design space for improvements.}~\label{fig:imls}
\end{figure}
Recommender systems are a core component of many intelligent and personalized services, such as e-commerce destinations, news content hubs, voice-activated digital assistants, and movie streaming portals. Figure~\ref{fig:imls} shows the high-level  structure of a typical recommender system. People interact with the recommender system via a front-end graphical user interface (GUI), which serves at least two roles. First, the front-end presents the predictions that the machine learner makes (e.g., personalized news stream, interesting items). Second, the front-end relays back any feedback data (e.g., clicks, ratings, or other signals stemming from user interactions) to the machine learner in the back-end. It is this aspect of feedback generation that we will explore in this paper.

Feedback data is typically categorized as either \emph{explicit}, where the user provides an explicit assessment like a star rating or a thumbs up, or as \emph{implicit}, where feedback is gleaned from how people use the system (e.g., navigation, dwell time) \cite{konstan1997grouplens}. We focus on implicit feedback in this paper, because its abundance and unobtrusiveness make it particularly attractive for the training of recommender systems. For example, clicked items are commonly used as positive learning examples, and the machine learner then predicts other items that the user would click on. In addition to user-specific data, the machine learner also typically has access to shared data, including interactions from other users, or other pre-collected information.

For a sustained positive user experience in recommender systems, the goal is to make accurate and personalized predictions that match a user's preferences. Machine learning algorithms do this by taking data from past user interactions as implicit feedback, and then generalize from the data in order to provide useful predictions in future interactions~\cite{Mitchell97}. In the vast majority of previous work, improving the accuracy of predictions has been approached purely from an algorithmic standpoint~\cite{konstan2012recommender}, e.g., by equipping the learning algorithms with more powerful generalization mechanisms such as deep learning~\cite{bengio2009learning}. While the algorithmic approach has led to the availability of a large zoo of specialized learning methods, gains in prediction performance are typically small and have theoretical limits. Another issue is that the algorithmic focus often ignores the processes that underlie the generation of feedback data, even though feedback data is a central ingredient to successful learning. Misinterpretation of the feedback data can substantially hurt learning performance~\cite{Joachims/etal/17a}, and failing to obtain abundant feedback data can render even the most sophisticated machine learning algorithm ineffective.

We argue that these problems arise because a purely algorithmic approach is a rather myopic strategy for optimizing learning effectiveness in recommender systems. As Figure~\ref{fig:imls} shows, recommender systems have several components beyond the learning algorithm, providing a rich design space for avoiding these problems. In this paper, we propose a more holistic, alternate way to improving learning effectiveness based on the insight that much of the effectiveness of the machine learner comes from feedback data itself. {\em Our core idea is to deliberately shape the implicit feedback data that is generated through the design of the user interface.} In particular, we explore how the design of the user interface can increase the quality and/or quantity of implicit feedback, which will render machine learning more effective in the long term, while maintaining a desirable user experience in the short term to avoid users abandoning the service that the recommender system is part of. 

To structure the space of interface designs with respect to their impact on implicit feedback generation, we rely on information foraging theory. Information foraging theory describes how people seek information in unknown environments. In particular, we explore \emph{foraging interventions} and their effect on implicit feedback along two directions -- the strength of the {\em information scent}, i.e., the amount of information that is displayed initially for an item, as well as the {\em information access cost} (IAC), i.e., the cost for obtaining more information about an item. We illustrate the effect of these interventions in a recommender system for the common task of selecting a movie to watch. 

This paper makes the following contributions:
\begin{itemize}
    \item We present the concept of foraging interventions for optimizing implicit feedback generation in recommender systems. We carry out a user study of these foraging interventions for the task of picking a movie to watch, showing that these interventions can effectively shape the implicit feedback data that is generated. Moreover, we find that the right foraging interventions not only help feedback generation, but that they also lower cognitive load.
    \item We propose to assess implicit feedback data along two dimensions that are important for machine learning effectiveness -- feedback quality and feedback quantity. Not only are these two dimensions intimately tied to statistical learning theory, but they also provide a layer of abstraction that makes them applicable to a wide variety of tasks and settings. Furthermore, we provide concrete metrics for assessing feedback quality and quantity and demonstrate their usefulness in our user study.
    \item On a conceptual level that goes beyond our user study, this paper establishes a more holistic view on improving recommender systems, recognizing that the same goals (e.g.\ improved predictions) can be achieved through different and complementary means (e.g.\ interface vs.\ algorithmic improvements). We ultimately hope that this will foster closer collaboration of machine learning specialists and UX designers, and we discuss possible areas of advancement.
\end{itemize}

\section{Related Research}
The research in this paper relies on prior work in various areas, specifically information foraging from cognitive psychology, learning from implicit feedback from machine learning, and interface design and evaluation from human-computer interaction.

\subsection{Information Foraging}
Informaging foraging theory \cite{pirolli1999information} is a theoretical framework originally proposed by Pirolli and Card in the 1990s that models how people seek information. Information foraging theory has its roots in optimal foraging theory~\cite{charnov1976optimal} that biologists and ecologists developed in order to explain how animals find food. Common to these foraging theories is that they are based on the observation that people or animals tend toward rational strategies that maximize their energy or information intake over a given time span. 
In information foraging, people called informavores or predidators hunt for prey (pieces of valuable information) that is located on different information patches in a topology (all patches and links between them). To locate prey, informavores need to continuously evaluate cues from the environment and can either (i) decide to follow a cue or (ii) move on to another information patch. The decision of which cue to follow is based on its information scent, i.e., an informavore's estimate of the cost and the value of the additional information that could be obtained by following the cue.

Information foraging theory was originally developed to explain human browsing behavior on the web~\cite{chi2001using,pirolli2007information}. It not only has served as a base for sophisticated computational cognitive models of user behavior \cite{fu2007snif,teo2012cogtool}, but also has had an impact on practical guidelines for web design \cite{larson1998web,spool2004designing}.
Moreover, it inspired the design of various user-interfaces and information gathering systems that try to assist users in 
certain steps during browsing, such as information scent evaluation~\cite{wexelblat1999footprints,olston2003scenttrails,fu2010facilitating}. 

A crucial component in information foraging theory is the concept of opportunity cost since it determines when an informavore moves on to the next information patch. Opportunity cost, or information access cost (IAC), can have important consequences for the actions and strategies that informavores employ. For example, it is known that people prefer extra motor effort over cognitive effort \cite{gray2006soft}. 
As an even more concrete example, it has been observed that this makes people more likely to memorize items with a button click in recommender systems, rather than memorizing the items themselves \cite{schnabel2016using}. There are also more specialized cost-benefit models \cite{azzopardi2016analysis}. In our study, we vary both the strength of information scent as well as the information access cost (IAC). 

\subsection{Implicit Feedback in Machine Learning}
As mentioned before, abundant and high-quality feedback data is essential for interactive systems that learn over time. Since implicit feedback is both readily available as well as unintrusive, it has been a main driver in the development of better interactive systems, such as information retrieval systems \cite{claypool2001implicit,fox2005evaluating,joachims2007evaluating,radlinski2008does}, recommender systems \cite{hu2008collaborative,rendle2009bpr,koren2009matrix}, and voice-activated intelligent assistants \cite{jiang2015automatic,kiseleva2016understanding}.  
Since implicit feedback data reveals people's preferences only indirectly, much research has focused on studying the quality of feedback signals that arise in various settings, i.e., studying how well implicit feedback reflects explicit user interest. Conventional mouse clicks often reflect preferences among search results \cite{radlinski2008does}, and mouse hovers are a good proxy for gaze \cite{huang2011no}. There has also been some research on alternative signals, such as cursor movements \cite{huang2012user,huang2011no,zen2016mouse} or scrolling \cite{guo2012beyond,liu2017scroll}. In our work, we ask the question of how we can shape this implicit feedback using foraging interventions in the systems interface, rather than studying how implicit feedback signals arise in fixed interfaces.

\subsection{Interface Design for Machine Learning}
With Machine Learning being an essential component in many interactive systems, UX researchers have started looking into Machine Learning as a design material~\cite{dove2017ux,mlenhanced2017yang}. However, there is relatively little work in the intersection of Machine Learning and UX design. Traditionally, since explicit feedback data has a longer standing in Machine Learning, most research in the intersection of UX design and Machine Learning also focuses on improving explicit feedback elicitation.
For example, Sampath et al.~\cite{alagarai2014cognitively} show that different visual designs can alleviate some of the cognitive and perceptual burden of crowd workers in a text extraction task which consequently increases the accuracy of their responses. Another example is the work of Kulesza et al.~\cite{kulesza2014structured}, where the authors present an interface that supports users in concept labeling tasks that are typical in Machine Learning. The interface supports users by allowing for concepts to be adapted dynamically in various ways during a task. Implicit feedback has received less attention, but recently has increased in importance \cite{dove2017ux,mlenhanced2017yang}. As a special purpose application, Zimmerman et al.~\cite{zimmerman2007vio} present an information encoding system that was designed to offer suggestions that improve through implicit feedback. Finally, Schnabel et al.~\cite{schnabel2016using} show in recent work that by offering users digital memory during choice tasks, one can significantly improve recommendation accuracy.

\section{Goals and Case-Study Environment}
\begin{figure}[t!]

  \centerline{\includegraphics[width=1.05\columnwidth]{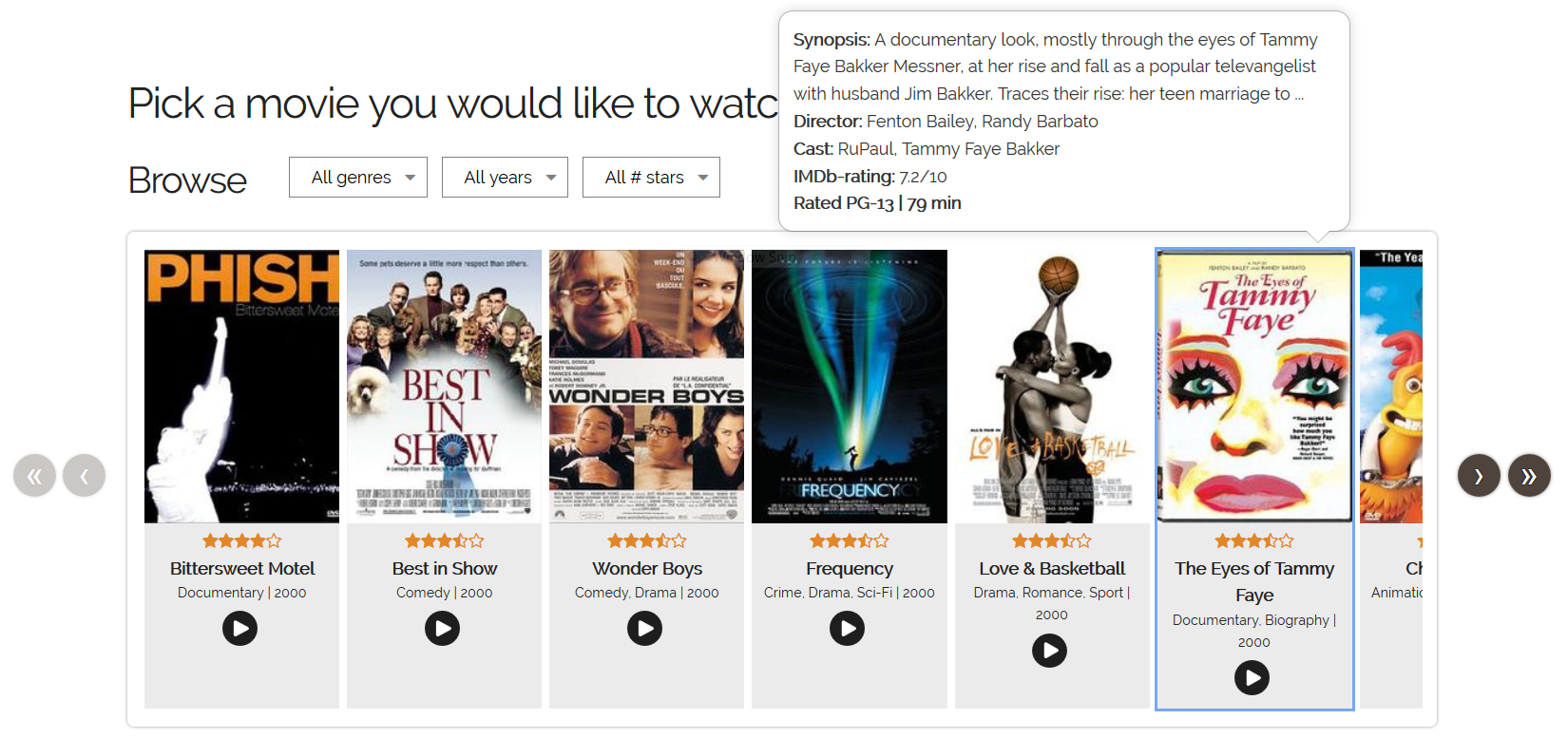}}
  \caption{Basic carousel interface for browsing movies that we used in our case study. The top right shows a popover that is displayed when a user requests more information about a movie.}~\label{fig:basic_interface}
\end{figure}
We illustrate the effects of foraging interventions in the context of a movie browsing interface in which we want to generate implicit feedback for session-based recommendation. Subjects in our study had the task of selecting a movie they would like to watch. Session-based recommendation is different from long-term recommendation, and it recognizes that users may have different preferences in different contexts. This means that session-based recommendation systems need to rely on implicit feedback from within the current session itself. 

The basic browsing interface is shown in Figure~\ref{fig:basic_interface}. We chose this interface because it mimics the interface of common online movie streaming services, such as Netflix, Hulu, or Amazon. Users can page through the list of movies using the arrow buttons, with extra buttons to jump to the beginning and end of the list. They can also filter the current view by genre, year, and OMDb star rating. To get more information about a movie, users can open a popover as shown for the right-most item in Figure~\ref{fig:basic_interface}. From the perspective of generating implicit feedback for learning the user's preferences for a particular session, we will use engagement with a movie -- specifically requesting the popover and the time reading the popover -- as an indicator of interest and preference. To eventually choose a movie to watch, users click the play button and confirm a final prompt.

In order to better understand the research hypotheses we are going to formulate next, we are going to analyze this interface through the lens of information foraging theory. In terms of information foraging theory, a carousel view can be seen as an information patch, and all possible carousel views form the topology. The goal of a user in the foraging loop \cite{pirolli2007information} is to collect enough information about the items in order to reach a decision. Captions and movie posters form cues that are associated with an information scent, and the effort and time that is needed to obtain more details about a movie constitute the information access cost. 

\newcommand{\specialcell}[2][c]{%
  \begin{tabular}[#1]{@{}c@{}}#2\end{tabular}}
\begin{table}[t]
\small
  \centering
    \begin{tabular}{rcc}
    \toprule
          & \multicolumn{2}{c}{information access cost (IAC)} \\
    \cmidrule{2-3} 
    scent & low   & high \\
    \midrule
    weak   & \specialcell{(only poster;\\hover for more)} & \specialcell{(only poster;\\click for more)} \\
    strong  & \specialcell{(title, year, rating, genre;\\hover for more)} & \specialcell{(title, year, rating, genre;\\click for more)} \\
    \bottomrule
    \end{tabular}%
  \caption{Crossed independent variables of our user study. Participants were exposed to an adjacent pair (row- or column-wise) of conditions.\label{tbl:conditions}}%
\end{table}%

The discussion above suggests two straightforward variables that can be manipulated by the interface designer -- the strength of the information scent, and the information access cost for viewing more details about a movie. We translate this into the following interface interventions for our recommendation setting.
\begin{description}
\item[Information Scent Intervention:] We map a weak information scent to a carousel interface in which only the poster images are shown; a strong information scent gets represented by adding a textual description to each movie that contains the year, genre and star rating of each movie (shown in Figure~\ref{fig:basic_interface}). 
\item[Information Access Cost Intervention:] We map low IAC to an interface where people simply hover over an item to display the popover (top right of Figure~\ref{fig:basic_interface}), whereas they need to click on an info button next to the play button in the high information access cost condition (not shown). 
\end{description}
In our study, we will look at all four possible combinations of these interventions, as summarized in Table~\ref{tbl:conditions}. Next, we will discuss which research questions we address in our user study.



\subsection{Research Hypotheses}
\label{sec:hypotheses}
The main focus of this paper is on understanding how foraging interventions drive the generation of feedback data, and we want to study the effects both in terms of the quantity and the quality of the feedback data. Moreover, we would like to understand how user experience and usability are affected -- especially in terms of cognitive load as well as overall preference. Below are the four research hypotheses that we formulated based on information foraging theory applied to our setting.

\begin{description}
\item[H1:] People will prefer more detailed descriptions over less detailed ones, or in other words, prefer the interface that provides stronger information scents.
\item[H2:] People will prefer having to hover instead of having to click for more information. In terms of information foraging theory, people seek to have low information access costs for obtaining more details.
\item[H3:] More detailed descriptions will also lead to increased feedback quality. Since the information scent is stronger, people are less likely to follow unhelpful cues.
\item[H4:] Feedback quantity will increase in cases when people can hover to obtain more information instead of having to click. Put differently, feedback quantity increases when information access cost decreases.
\end{description}

Note that H1 and H2 study foraging interventions through a HCI-based perspective since they only assess user outcome, whereas H3 and H4 focus on the consequences of foraging interventions for ML. Obviously, we would like the optimal interventions (e.g., strong scent, low IAC) to be the same from both the HCI as well as ML perspective, but this is not guaranteed.

\section{Study}
To test our hypotheses outlined above, we ran a crowd-sourced user study where we varied both information access cost and information scent in the browsing interface of Figure~\ref{fig:basic_interface}.

\subsection{Design}
We used a repeated measures mixed within-and between-subjects design with the following two independent variables: information scent (weak, strong) and information access cost (low, high). Each subject conducted two movie-selection sessions. This allowed us to have a sensitive within-participant design for the variable of interest that could be completed within a reasonable amount of time.
Table~\ref{tbl:conditions} shows the four combinations resulting from a fully crossed design. For the mixed design, we varied one of the two variables between subjects (e.g., information scent) and the other one within subjects (e.g., information access cost). For example, a participant would be assigned to a condition where the information scent was held fixed, but the the information access cost was varied between the two movie-selection sessions. In total, there were four such pairwise conditions, each corresponding to an adjacent pair (row- or column-wise) in Table~\ref{tbl:conditions}.

In each study, the order in which the interventions were displayed was counterbalanced, as well as the partition of movies that was displayed. We had a total of 2310 movies that was divided into two non-overlapping partitions of 1155 movies each in order to have a fresh set of movies in each session. This ensured that in the second task the participant was not subject to a learning effect over the set of movies in consideration.

\subsection{Participants}
We recruited 578 participants from Amazon Mechanical Turk. We required them to be from a US-based location and have at least a task approval rate of 95\%. To ensure that the display contents were as similar as possible, we required a minimum screen resolution of $1024\times768$ px and a recent compatible web browser. Moreover, the movie data needed by the interface was cached locally before the study started, so that response times were independent of the speed of the internet connection.

Participants were paid \$1.80 upon the completion of our study. With an average completion time of about 10 minutes for the study, this resulted in an effective wage of \$10.80 an hour which is well above the US Federal minimum wage. 
The mean age was 34.4 years (SD=9.5), with 61\% of the participants being male, and 39\% female. Most participants (42\%)  reported that they would watch a movie more than once a week, followed by 37\% of people watching movies less often than that but at least once a month. 15\% of people said they watched a movie on a daily basis, and the remaining 6\% of the participants said they would watch a movie less than once a month.

\subsection{Procedure}
\label{sec:procedure}
At the beginning, users were informed about the basic structure of the study. We gave them the following prompt:
\begin{displayquote}
Your task is to choose a movie you would like to watch \emph{now}. 
\end{displayquote}
We chose this task prompt because it reflects a common task in the recommender systems literature, called the \emph{Find Good Items} task  \cite{herlocker2004evaluating}. After viewing the instructions, they were assigned randomly to a condition and were presented with a brief information graphic that explained how they could obtain more information about an item (either by hovering or by clicking on the information button). 
Participants then had to choose a movie using the first interface from the first partition of movies. After that, they were reminded of the task prompt before having to select a movie again using the second interface and the second partition of movies. At the end, participants were asked to fill out a survey about their experiences with the two interfaces. Moreover, participants had to rate some of the movies with which they interacted previously. 

\subsection{Measurements}
We measured usability in terms of user performance as well as subjective ratings. Measurables for user performance included: task completion time, number of navigate operations, number of items displayed. For subjective ratings we elicited overall interface preferences, and to measure cognitive workload, we used the NASA Task Load Index (TLX)~\cite{hart1988development}. 
We also logged hover events as well as all events associated with opening and closing a popover to track the quantity of the generated implicit itemwise feedback data. Finally, users were asked to rate how much they would like to see certain movies right now on a 7-point Likert scale. This was done in order to be able to study implicit feedback quality (e.g., how hover time relates to Likert rating). 

\subsection{Movie Inventory}
To make our system similar to real-world movie streaming websites, we populated it with data from OMDb\footnote{http://omdbapi.com/}, a free and open movie database. We only kept movies that were released after 1950 to ensure a general level of familiarity. In order to have a reasonably attractive default ranking, we sorted all movies first by year in descending order, and then by review score (OMDb score), again in descending order. This means that people were shown recent and highly-rated movies first in the browsing panel by default. Plot summaries of movies were shortened if they exceeded two sentences to make all synopses similar in length.

\section{Findings}
We now present the results of our case study, focusing first on the subjective measurements, and then on the quantitative and qualitative aspects of the implicit feedback data that has been generated. All significance tests in this section use a p-value of $p < 0.05$.
Overall, we had 578 users complete the study. According to current best-practice recommendations for quality management of crowdsourced data~\cite{kittur2008crowdsourcing,komarov2013crowdsourcing}, we  used a mix of both outlier removal as well as verifiable questions. More concretely, we employed the following criteria for filtering out invalid user responses by removing:
\begin{compactitem}
   \item Users that spent less than 15 seconds in each session;
   \item Users that were inactive for more than 90 seconds in a session;
   \item Users that reloaded the interface page;
   \item Users that rated their final chosen item at 3 points or lower.
\end{compactitem}
In the end, we were left with 444 valid user responses, with each condition having between 107 and 114 responses.

\subsection{Interface Usability and User Preference}
\begin{figure}
\centering
  \includegraphics[width=0.97\columnwidth]{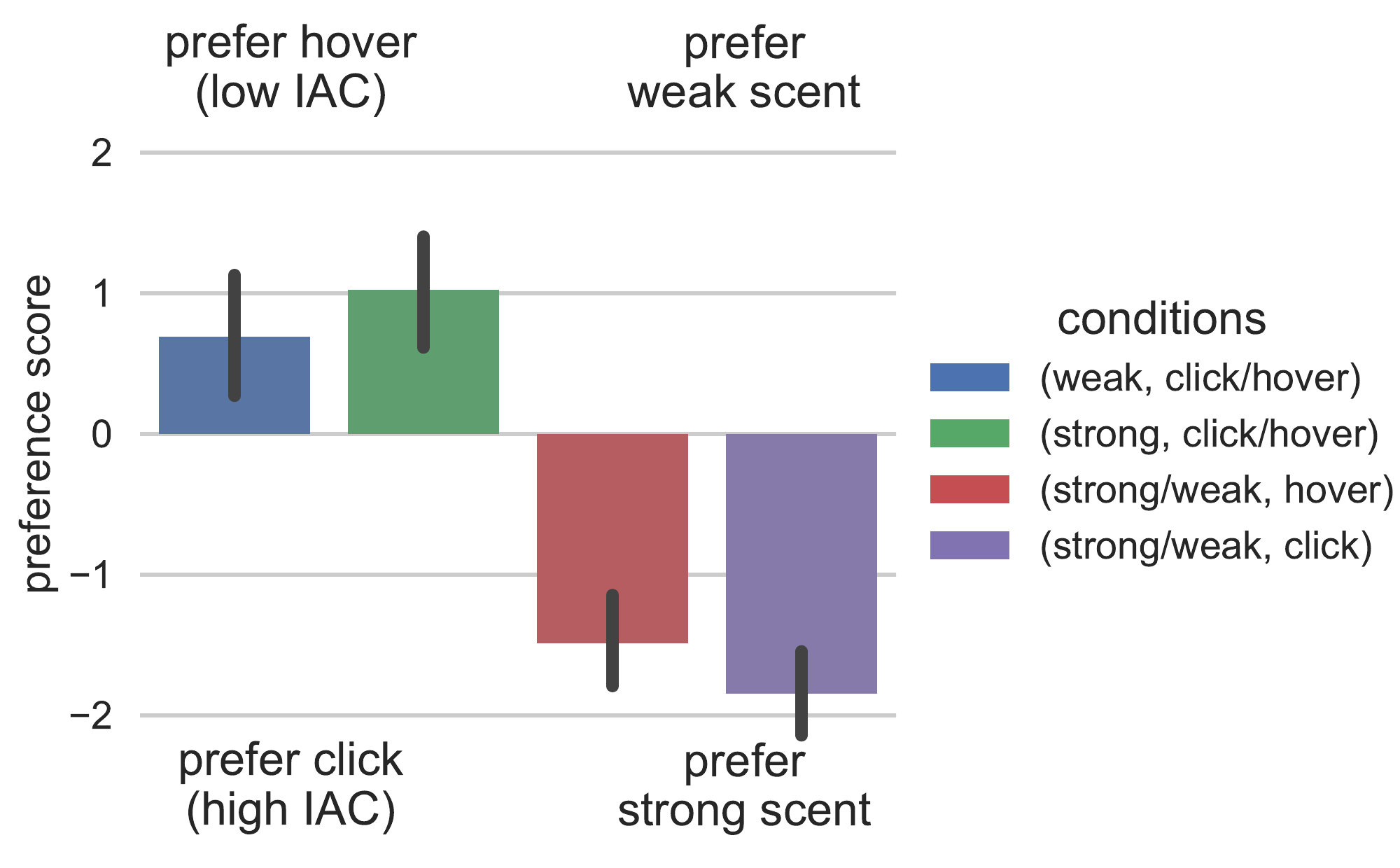}
  \caption{Overall preference for interfaces in paired setup.}~\label{fig:overall_preference}
\end{figure}

In the final survey, we asked people for their overall preference between the two interfaces with which they interacted. Overall preference was measured on a 7-point scale ranging from a strong preference for the first interface (-3) over neutral (0) to a strong preference for the second interface (+3).
Figure~\ref{fig:overall_preference} shows the mean preference score for each condition. A positive score means that the intervention after the slash (see legend) was preferred, whereas a negative score means an overall preference for the intervention listed before the slash. 
As we can see from the figure, people significantly prefer to hover for more information than to click (t-test; different from zero). Likewise, we can see that in the conditions where we compared a weak to a strong information scent, people significantly prefer the interface with strong information scent over the interface with a weak one. Interestingly, preferences appear to have a stronger magnitude in the conditions where information scent was varied (cf. the two rightmost bars in Figure~\ref{fig:overall_preference}).
\begin{figure*}
\centering
  \includegraphics[width=0.97\linewidth]{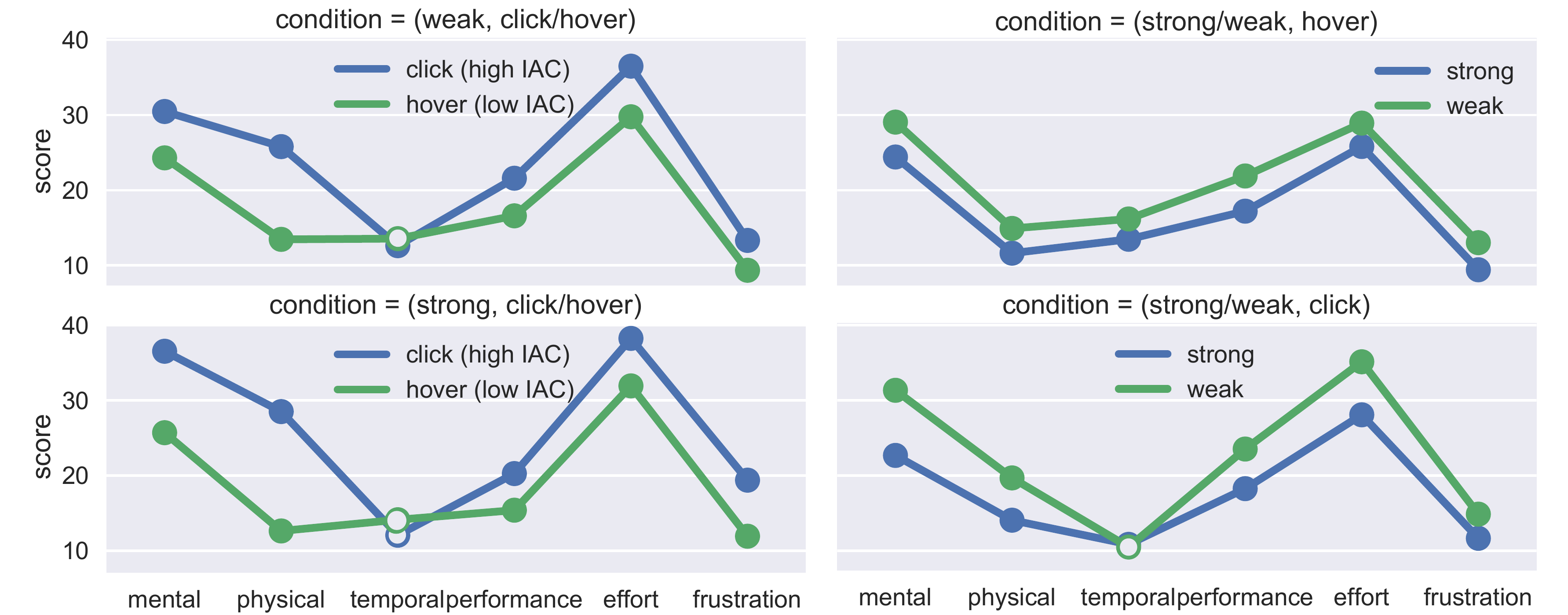}
  \caption{Individual responses to NasaTLX, lower scores meaning better. Lines were added to enhance readability. Solid circles indicate statistical significance.}~\label{fig:nasa_tlx}
\end{figure*}

We also measured cognitive load using the NasaTLX questionnaire~\cite{hart1988development}. Figure~\ref{fig:nasa_tlx} shows the raw scores for each of the subscales with lower values meaning better. We used solid circles whenever the pair of subconditions has significantly different scores, and hollow circles otherwise (pairwise t-tests). Looking at the left column first, we can see that NasaTLX scores were significantly lower for the hover interface on all subscales but on the temporal demand one. The difference in TLX scores is largest for the physical demand subscale where people were asked to rate how much physical activity was required during the task.
The right column in Figure~\ref{fig:nasa_tlx} shows user responses for the conditions where information scent was varied. Similar to overall preference, the strong information scent interface achieves better TLX scores on all subscales except on the temporal demand one. Differences in scores seem to be more pronounced when people had to click to obtain more information.
Another interesting finding is that in all conditions, the change in interface also affected how successful people perceived they were at the task (cf. the performance subscale).

We also asked people to explain briefly why they preferred one interface over the other. We grouped their answers manually according to which property they focused on. For the two conditions where we varied the information access cost, most people mentioned that hovering provided a more intuitive way of accessing more information (30\% total), followed by 27\% of people that stated that they found hovering to be less effort. Among the people that preferred clicking for more information instead of  hovering, the most frequently reported reason was that this allowed them to focus on one item at a time (13\%).
In the low vs.\ strong information scent conditions, people frequently mentioned they enjoyed having the star rating (35\%) as well as more information in general with each movie (34\%). The most frequent reason for people to prefer the interface with a lower information scent was that it felt cleaner and less cluttered (5\%).



\subsection{Feedback Quality}

\begin{figure}
\centering
  \includegraphics[width=0.97\linewidth]{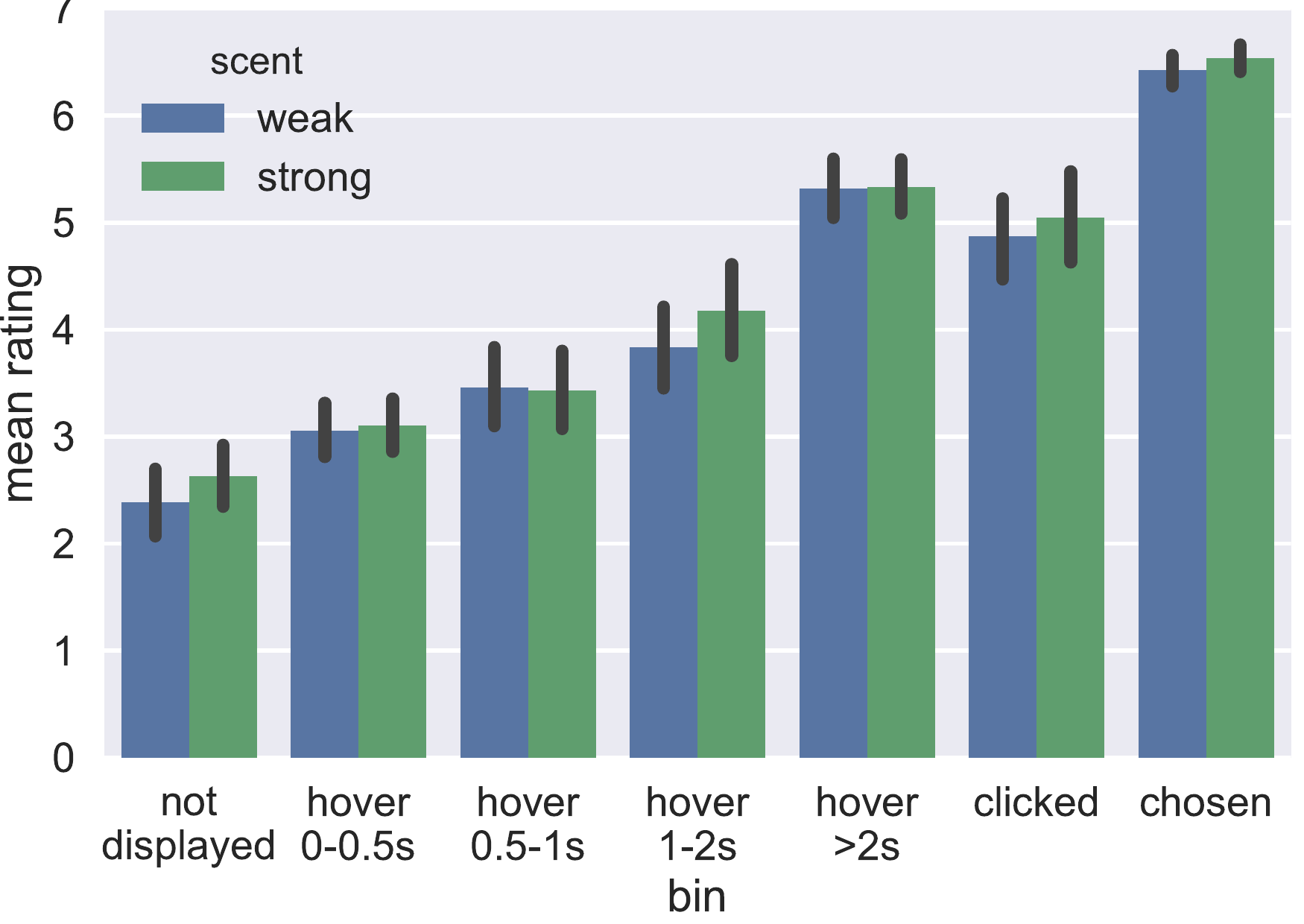}
  \caption{User ratings for items in different interaction categories. The condition is (strong/weak, click).}~\label{fig:feedback_quality}
\end{figure}
Statistical Learning Theory \cite{shalev2014understanding} and practical evidence (e.g., \cite{basilico2004unifying}) exhaustively shows Machine Learning algorithms become better with increasing data quality (lower noise) and quantity (number of training samples). Motivated by this, we use feedback data quality and quantity in our study as key metrics for assessing learning effectiveness. In this section, we first study how feedback quality changes under our interventions, before turning to feedback quantity in the next section. We define feedback quality here as how well observable user interactions reflect a preference for an item. Feedback quality was analyzed by aggregating user ratings for observable user interactions from different categories. More concretely, to assess the quality of feedback signals, we propose to aggregate the explicit user ratings that were elicited in the post study questionnaire, where we asked people how much they would like to watch particular movies now on a 7-point scale from not at all (1) to very much (7). We asked for ratings for movies coming from seven different interaction categories. The first category contained all movies that were never displayed to the user. Then came four categories that sampled movies with different hover times.
We also had an extra category for movies that were clicked on in the "click" condition for the popover, and then a final category for the movie that was chosen by the user in the end (even though the latter feedback signal comes too late for session-based recommendation).

Figure~\ref{fig:feedback_quality} shows the average user ratings for each category. To compute the mean rating, we first averaged all ratings from a single user that were sampled uniformly from a category so that each user contributed equally, and then averaged this mean value across all users. We can see that items that were never displayed to the user as well as items that were hovered less than 500 ms rank lowest in terms of average user rating. Increasing the threshold to 500-1000 ms tends to increase the mean rating slightly, items with hovers of 1000-2000 ms length are rated with 4 points (middle of the scale) on average. The first category with an obvious interest are long hovers (2000 ms and more), followed by items that were clicked on. There are no significant differences between the longer hovers and clicked items average rating. Naturally, items that were participants' final choices have the highest average rating which comes in just slightly below the maximum rating of 7.

Figure~\ref{fig:feedback_quality} only shows the results of one of the pairwise conditions, namely (strong/weak, click). The error bars correspond to the standard error. As we can see from their magnitudes, there are no significant differences between the average ratings of the bins between sessions where information access cost was varied. Even though not shown here, this observation is the same for all three remaining conditions. 

From Figure~\ref{fig:feedback_quality}, we can also see that feedback from two categories is of especially high quality -- namely clicked items and items with long hovers. These are the events that are most interesting from a Machine Learning perspective since they reflect user interest well and can be used to define positive examples to the learning algorithm. We hence focus on quantity of events within these two categories in the following section. 

\subsection{Feedback Quantity}
The subjective measurements above indicate consistent user preferences for high information scent, independent of the information access cost, as well as an overall preference for low information access cost, independent of the information scent. We now address the question to which degree our interventions shaped the quantity of feedback data. To start with, we did not find any differences in the time-to-decision or in the number of browsed items between the interfaces in any of the four conditions.

\subsubsection{High vs. low information access cost}
\begin{figure}
\centering
  \includegraphics[width=0.97\linewidth]{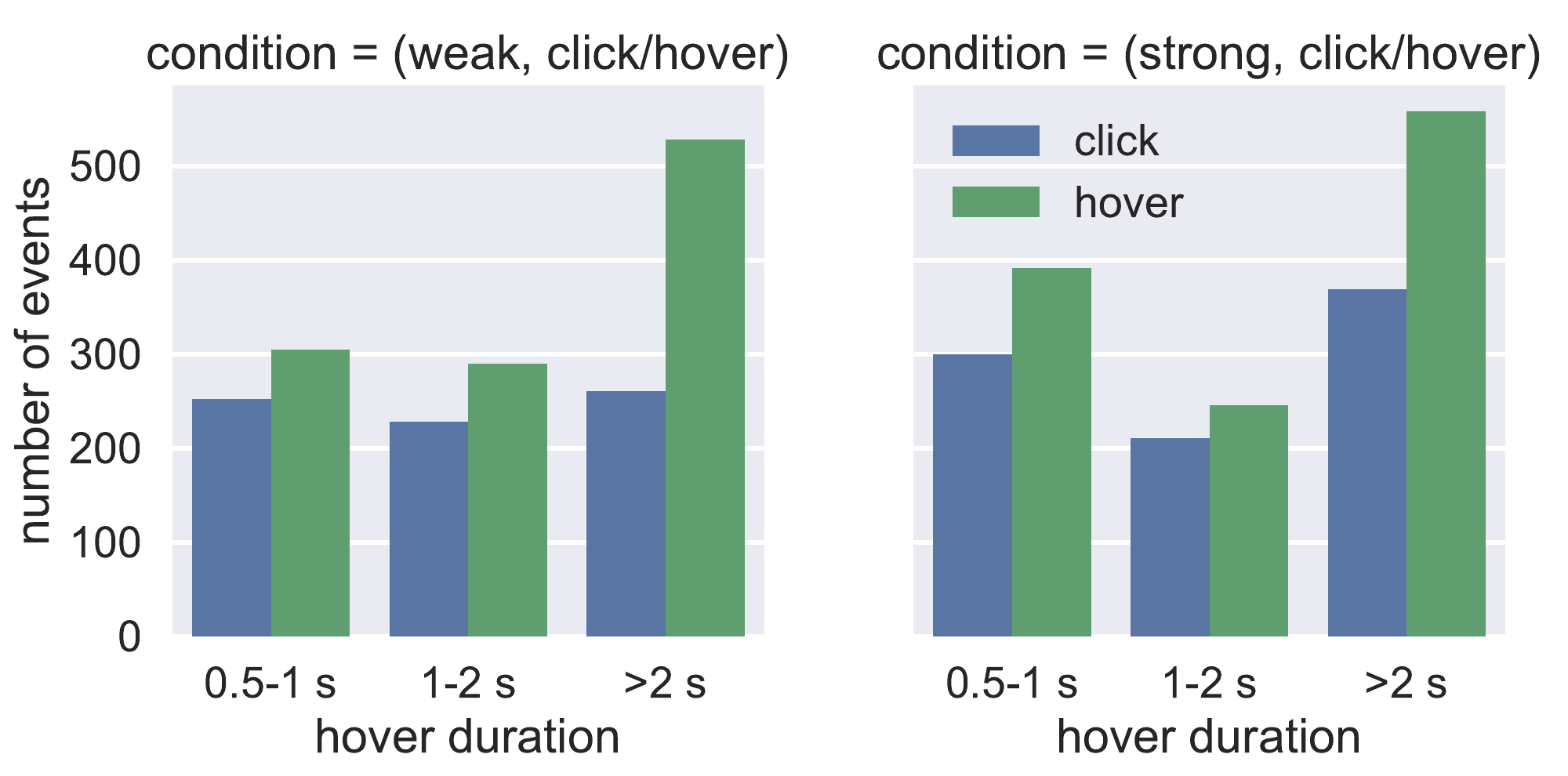}
  \caption{Hover events split by duration. By lowering the IAC, the number of long hovers increases.}~\label{fig:click_hover_quantity}
\end{figure}

\begin{table}[t]
  \centering
    \begin{tabular}{rrrr}
    \toprule
     \phantom{abcdefgh}  &   & \multicolumn{2}{c}{interface} \\
    \cmidrule{3-4} 
    &scent & click & hover \\
    \midrule
    \multicolumn{4}{l}{\emph{hover events > 500 ms:}} \\
    &weak  & 6.92  & 10.49* \\
    & strong & 7.71  & 10.47* \\
   \midrule
   \multicolumn{4}{l}{\emph{positive feedback events:}} \\
   & weak  & 3.34  & 4.93* \\
   &  strong & 	3.59 &	4.89* \\
   \bottomrule
    \end{tabular}%
  \caption{Feedback quantity under varying information access cost. Lowering IAC significantly increases feedback quantity. \label{tbl:hover_iac}}
\end{table}%

We first turn to the effect of information access cost on feedback quantity. In our mixed design, this means that we only analyze user responses by fixing information scent and varying information access cost. Table~\ref{tbl:hover_iac} presents the mean number of hover events that exceed 500 ms in each condition. We analyze our mixed design within subject which corresponds to column differences in Table~\ref{tbl:hover_iac}. Row differences are between-subject and higher variance. Looking at the first row, one can see that the low IAC condition produced roughly 54\% more short hover events (paired t-test) than the high IAC condition in the context of a weak information scent. A significant increase in short hover events can be observed in the context of a strong information scent as well, albeit the magnitude is somewhat smaller -- the number of events rises only by 26\%. Note that using hovers as an implicit feedback signal makes sense even when hovering is not connected directly to an action in the click condition, since hovers are known to track gaze while reading for many users \cite{huang2011no}. 

Figure~\ref{fig:click_hover_quantity} shows the number of hover events split by duration. We can see that increases in the number of hovers occur in all bins, but especially longer hovers (over 2000 ms) increase when hovering triggers the popover. Compared to the strong scent condition, the increases are again larger in the context of the weak information scent condition where the number of long hover events almost doubles. However, even in the context of the strong information scent condition, the number of long hover events still increases by about 50\%. 

Hover events are not the only way that users can express interest in an item -- in the click setup they can also get information by pressing the corresponding information button. This suggests that we should consider both long hovers as well as clicks as positive feedback. The bottom of Table~\ref{tbl:hover_iac} reports the mean number of positive feedback events where a positive event is either an explicit click or a longer hover (2000 ms or longer). Independent of the strength of information scent, we see significant increases in the number of positive events (paired t-test). The increase is slightly larger when items have a weak scent when compared to the increase under a strong information scent --- 32\% vs. 27\%.

\subsubsection{Strong vs.\ weak information scent}
\begin{figure}
\centering
  \includegraphics[width=0.65\linewidth]{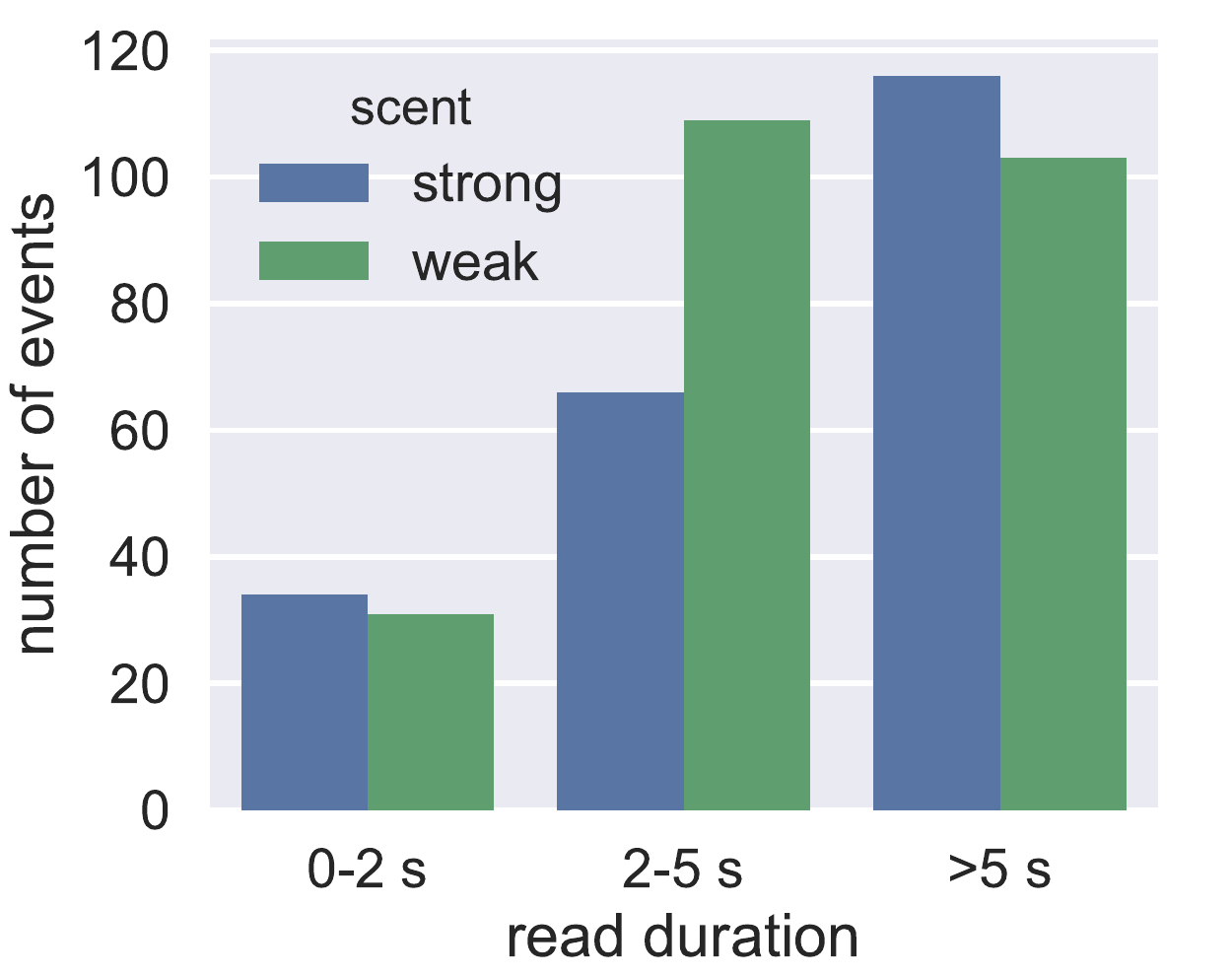}
  \caption{Read events split by duration. In the weak information scent setting, medium-length reading events increase.}~\label{fig:reading_times}
\end{figure}
Next, we turn to the influence of the strength of the information scent on feedback quantity. We now look only at user responses of the conditions where we fixed information access cost and varied the information scent by showing additional information with each item. Table~\ref{tbl:quantity_is} shows the overall number of hover events in each condition. If the interface only required hovering for more information, we can see that there is a significant increase in the number of hovering events when the information scent gets weaker. We can observe a similar trend in the click condition, however, the increase is not statistically significant. 

\begin{figure}
\centering
  \includegraphics[width=0.97\linewidth]{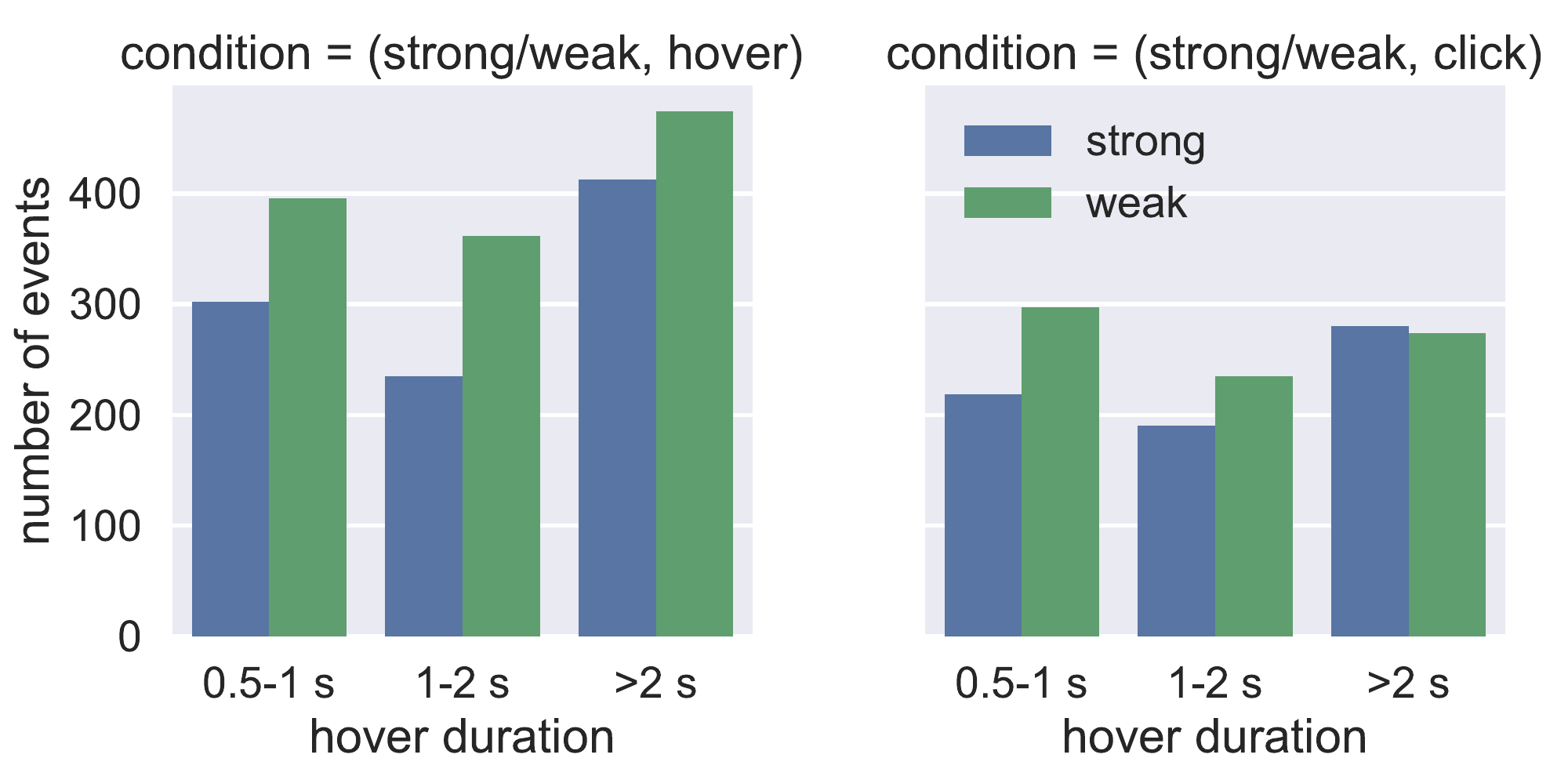}
  \caption{Hover events split by duration. Blue bars correspond to events under strong information scent interface; green bars to events under weak scent.}~\label{fig:weak_strong_quantity}
\end{figure}

Figure~\ref{fig:weak_strong_quantity} presents again a more fine-grained analysis of the hover events binned by duration. Most noticeable is that in both conditions, independent of the information access cost, there is no significant increase in the number of long hover events. However, we did find significant differences in the lowest bin (500-1000 ms) for both conditions, as well as in the middle bin (1000-2000 ms) for the condition where people used hovering. This also prompted us to look at the distribution of reading times in the condition where users could only obtain more information by clicking. Figure~\ref{fig:reading_times} shows the number of reading events binned by duration. Interestingly, there is also a significant effect in the number of medium-length read events (2000-5000 ms) which almost doubles when making the information scent weaker.

Looking at the union of positive feedback events, we also failed to find significant differences between sessions with weak and strong information scent. Overall, this means that the impact for information scent on overall positive feedback event quantity is weak or non-existent in our experiments. However, we did observe differences in the number of short and medium-length hover events.

\begin{table}[t]
  \centering
    \begin{tabular}{rrrr}
    \toprule
     \phantom{abcdefgh}  &   & \multicolumn{2}{c}{scent} \\
    \cmidrule{3-4} 
    &cost & strong & weak \\
    \midrule
    \multicolumn{4}{l}{\emph{hover events > 500 ms:}} \\
    &hover  &8.55 &	11.09* \\
    &click & 	6.14 &	7.20\phantom{*}  \\
   \midrule
   \multicolumn{4}{l}{\emph{positive feedback events:}} \\
   & hover  &3.72 &	4.27\phantom{*} \\
   &  click & 	2.88 &	3.08\phantom{*} \\
   \bottomrule
    \end{tabular}%
  \caption{Feedback quantity under varying information scent within-user strength. Feedback quantity increases mildly to moderately under weak scent.\label{tbl:quantity_is}}
\end{table}%

\subsection{Summary of Findings in Case-Study}
Our main findings can be summarized as follows:
\begin{itemize}
\item Overall people preferred the hover interface over the click interface, independent of the strength of the information scent.
\item People preferred the interfaces that provided a strong information scent over the ones with weak information scent.
\item Feedback quality remained largely unaffected by our interventions as measured by the average user rating per interaction bin. 
\item Feedback quantity measured as the number of positive events increased with decreasing information access cost. Varying information scent had a weaker effect on feedback quantity; it mostly shaped the distribution of hover durations.
\end{itemize}

\section{Discussion}
We now return to the research hypotheses we formulated earlier in Section~\ref{sec:hypotheses}.

Regarding the first research hypothesis, we did indeed find clear evidence that on average, people will prefer more detailed descriptions of items over less detailed ones.  This was true for both levels of information access cost. We also saw decreased cognitive load scores for almost all subscales of the NasaTLX. This is also in line with what we would expect using information foraging theory -- stronger information scent allows people to weed out non-promising leads only as well as make it less likely that they will miss out on a promising lead because of insufficient information. From the textual responses, it seems that the most important criterion for people in the movie domain was the star rating; this confirms the prominent position of star ratings that has been extensively studied in other research \cite{moe2011value,analytis2017social}. However, there were also some people opposing the stronger information scent, mainly because they felt that the interface became too cluttered. In the light of information foraging theory this can be explained by the increased cost for perceiving the information scent offsetting the benefits that came from a stronger scent.

We also found conclusive evidence that allows us to confirm the second research hypothesis. People overall preferred the interface with a lower information access cost, independent of the strength of the initial information scent. This was also reflected in overall lower NasaTLX scores. Information foraging theory delivers a simple explanation for this -- since an informavore's goal is to maximize their information intake under some constraints, lowering the information access cost should make it more efficient and hence more satisfied with its performance. 
This is also in line with with the qualitative responses that people provided -- most of them thought that hovering was more intuitive and less effortful. Again, there were also some people who disliked the interface with a lower information access cost, mostly because they felt it was hard to focus on single items with hovering, and that hovering made the interface more cluttered. Combined with the results of the first research hypothesis, this suggests that there is no one-size-fits-all interface since for some people the cost for processing additional information is higher than the benefit they get from more and easier information access.

Moving on to our third research hypothesis, we did not find direct effects on feedback quality as measured by the mean user rating in different interaction bins. This is despite the fact that people strongly preferred the interface that provided more information upfront. This could of course be a result of the way we operationalized feedback quality; and other notions would indeed show differences. However, there seems to be an indirect effect on quality through information scent affecting feedback quantities. For example, a weaker information scent led to an increase in short hovers -- probably to peek at items' descriptions. The good news is that for learning algorithms, time spent on an item correlates well with average user rating, independent of the foraging interventions we tried. This should make duration a stable signal overall for learning.

Regarding our last research hypothesis, we found consistent increases in feedback quantity under decreasing information access cost. Focusing especially on the category of highly indicative events such as long hovers or clicks, we found increases of at least 27\% in feedback quantity when people could hover instead of click to obtain more information. This finding also makes sense through the lens of information foraging theory -- assuming that people have a constrained budget, a reduced information access cost should enable them to consume more information, thereby increasing feedback quantity. 

Considering all foraging factors, our results suggest that from an HCI perspective (user preference and cognitive load) the optimal design point is strong information scent and low information access cost, but from a Machine Learning perspective there are advantages to weak information scent and low information access cost. 
This is because, looking at the within-user differences, we get an increase in total positive feedback events and a statistically significant increase in hover events  as Table~\ref{tbl:quantity_is} indicates.
However, the improvement in total quantity of hover events decreases as the IAC increases; we find that significance disappears and the gap shrinks. These results also imply that considering information access cost and information scent independently is not enough, because these factors can interact. 
Our results also highlight the importance of having low information access cost in interfaces. Although we only considered hovering vs. clicking as IAC interventions, we believe that low IAC is beneficial in many other settings as well. For example, news stories that use "click for more" to track attention better may want to consider prefetching and hovering to reduce IAC.

\section{Limitations}
An obvious limitation of the study in this paper is that people were paid for the completion of the experiment. Since this might change user behavior in various ways, for example shorten the overall time-to-decision, it would be interesting to repeat the experiment in an open setting to observe natural user behavior. Another interesting question is that of the strength of domain-dependent effects; since picking a movie is a quite visual task, it would be interesting to study domains, such as the task of having to pick a book where less information can be inferred from the pictorial representation of an item. 

\section{Future Work}
We believe that the approach that we took to improving recommender systems beyond the algorithm is just one of many viable interventions. Realizing that the recommender system is made out of many components (cf. Figure~\ref{fig:imls}) opens up a rich design space that should be explored future research. 

{\bf Better feedback regimes.} As we have argued in this paper, focusing only on the learning algorithm itself is myopic and limits our ability to improve the overall system. In this paper, we showed that by focusing on \emph{how} we get humans in the loop we can effectively shape the feedback data that gets fed to the learning algorithm. However, the interventions we studied in this paper are just the first step to developing a larger set of design patterns and interventions that we can use to shape or even design the feedback data that is needed for successful Machine Learning. In particular, it would be interesting to go beyond the typical binary feedback data regimes (e.g., relevant / not relevant), presenting users with more effective feedback mechanisms, where feedback also has a stronger semantic value. This also can then fuel the creation of new Machine Learning algorithms that are able to incorporate this kind of feedback data. An example for feedback regimes in this direction are faceting and filtering techniques~\cite{koren2008personalized,loepp2015blended}, enabling users to give set-based feedback while browsing.  Another example that shows how one can elicit a different type of feedback is \emph{shortlists}~\cite{schnabel2016using}. Shortlists provide an explicit mechanism for tracking all items that a user is currently interested in. Not only were shortlists able to increase user satisfaction, but also doubled the amount of feedback data in sessions where they were present. 

{\bf Better design processes.}
If we want to build better recommender systems, we need to also be more inclusive of Machine Learning goals during UX and UI design processes, with metrics that tell designers whether and how the ML performance is affected by certain design decisions. This is because similar to the algorithmic standpoint, focusing only on the UX design and being oblivious to the learning algorithms in the back-end has been argued to be myopic \cite{brown2016interaction,mlenhanced2017yang,dove2017ux}. The solution to this is to make Machine Learning also a design material in the UX design process~\cite{dove2017ux}. An obvious challenge is to make Machine Learning effectiveness a graspable concept to UX designers, with the right balance of abstractness as well as technical soundness. Our notions of feedback quality and quantity as surrogate measures to learning effectiveness could offer a starting point, and these can also be easily obtained in the user testing stage. 
The notion of feedback quality we presented in this paper only allows for a relatively course-grained analysis by looking at the mean user ratings in separate bins, so it would be interesting to explore preferential statements as an alternative that could perhaps produce a more detailed picture -- for example by ranking items that users interacted with in their session. It would also be interesting to look at other relevant properties of feedback data, such as how well it covers the entire space of items, to further enhance learning performance.
Overall, we are convinced that including factors like feedback quality and quantity into user studies offers new possibilities to improving UX in the long run, going beyond short-term usability and satisfaction. 

{\bf A better understanding of trade-offs.} To make more informed design decisions, we need to gain better insight into which and how different factors in a recommender system interact. A common trade-off is the one between long-term learning efficiency and short-term usability. For example, we found that hovering increased feedback quantity, but it was perceived to be intrusive by some users. With a better understanding of trade-offs, we might also be able to personalize some interface elements to account for different user types.



\section{Conclusions}
In this paper, we studied foraging interventions as an alternate and complementary way to making learning in recommender systems more effective. We conducted a user study of these foraging interventions for the common task of picking a movie to watch, showing that these interventions can effectively shape the implicit feedback data that is generated while maintaining the user experience. We introduced metrics for feedback quantity and feedback quality as surrogate measures for learning effectiveness, and argued that these offer an easy way for UX designers to incorporate Machine Learning goals into their design. Going beyond our case study of foraging interventions, this paper establishes a more holistic view on improving recommender systems, recognizing that there are many options for synergistic effects when building these systems. We ultimately hope that this will foster closer collaboration of Machine Learning specialists and UX designers.

\nottoggle{anonymous}{
\section*{Acknowledgments}
We thank Samir Passi and Saleema Amershi for their helpful comments on previous versions of this document. This work was supported in part through NSF Awards IIS-1247637, IIS-1513692, IIS-1615706, and a gift from Bloomberg.
}

\bibliographystyle{ACM-Reference-Format}
\bibliography{references} 

\end{document}